\RequirePackage[2020-02-02]{latexrelease}
\documentclass[twocolumn,amsmath,amssymb,prl,aps,showkeys]{revtex4}
\usepackage{graphicx}
\usepackage{amsmath,amssymb}
\usepackage{mathrsfs}
\usepackage{color}
\usepackage{afterpage}
\usepackage[version=3]{mhchem}
\usepackage{natbib}
\usepackage{soul}
\usepackage[caption=false]{subfig}
\usepackage{array}
\usepackage{multirow}
\usepackage{float}

\begin{document}

\title{Strain-Induced Modulation of Spin Splitting and Persistent Spin Textures in Low-Symmetry 2D Hybrid Perovskites: A case study of RP phase}

\author{Shantanu Pathak\footnote{shantanu.pathak@physics.iitd.ac.in}, Saswata Bhattacharya\footnote{saswata@physics.iitd.ac.in}} 
\affiliation{Department of Physics, Indian Institute of Technology Delhi, New Delhi 110016, India}
\begin{abstract}
	\noindent 
We report the observation of a persistent spin texture (PST) in pseudo-2D hybrid perovskite, characterized by significant spin splitting strength on the order of \(3 \, \text{eV} \cdot \text{Å}\). Using first-principles density functional theory (DFT) calculations, complemented by a \(\mathbf{k} \cdot \mathbf{p}\) model analysis, we validate the presence of PST and its robustness under various conditions. The material's non-centrosymmetric nature and strong spin-orbit coupling ensure uniform spin orientation in momentum space, enabling long spin lifetimes and promising spintronic applications. Furthermore, we demonstrate the tunability of the spin splitting via the application of external strain and stress, offering a versatile approach to control spin configurations. Our results highlight the potential of this perovskite system for next-generation spintronic devices, where external perturbations can be used to precisely modulate electronic properties.
 
\end{abstract}
\maketitle

\section{Introduction}

Hybrid organic–inorganic perovskites (HOIPs) have emerged as a fascinating class of materials due to their structural tunability and exceptional optoelectronic properties.\cite{zhang2020advances,zhang2022room,ren2021advances,stoumpos2016halide,pedesseau2014electronic} Initially popularized for solar cell applications owing to their high power conversion efficiencies (PCEs)—with CH$_3$NH$_3$PbI$_3$ (MAPbI$_3$) demonstrating a remarkable increase in PCE from $\sim$3.8\% in 2009 to over 25\% in 2020 \cite{kojima2009organometal,green2020solar}—HOIPs have also attracted interest in broader optoelectronic and quantum materials research. The stability and dimensionality of these systems are often governed by Goldschmidt’s tolerance factor,\cite{goldschmidt1926laws} where the incorporation of larger organic cations typically drives the structure toward lower-dimensional phases.\cite{tsai2016high,pedesseau2016advances,kovalenko2017properties} In particular, two-dimensional (2D) HOIPs, composed of alternating organic and inorganic layers, have gained traction due to their enhanced environmental stability and unique quantum-confined architecture. These naturally layered structures form multiple quantum well (MQW)-like configurations,\cite{liu2018tunable} in which the inorganic framework acts as a potential well and the bulky organic spacer serves as a tunneling barrier.

This geometry offers precise control over the electronic structure and spin–orbit coupling (SOC), tunable via chemical substitutions at the organic cation, inorganic lattice, or their interface. The presence of heavy elements such as Pb introduces strong SOC, lifting the spin degeneracy of conduction bands and giving rise to spin-split electronic states. This makes HOIPs a fertile platform for exploring spin-resolved physics, where the interplay among symmetry, SOC, and structural distortions enables the realization of exotic spin textures, including persistent spin textures (PSTs). As such, low-dimensional HOIPs stand out as promising candidates for next-generation spintronic and optoelectronic applications.

In non-magnetic systems, spin-degenerate bands are generally expected in the absence of a magnetic field. However, this degeneracy can be lifted in materials lacking spatial inversion symmetry due to SOC, which induces an effective momentum-dependent spin-orbit field (SOF) described by \(\boldsymbol{\Omega}(\mathbf{k}) \propto \mathbf{E} \times \mathbf{k}\), where \(\mathbf{E}\) is an internal electric field arising from inversion asymmetry \cite{dresselhaus1955spin,bychkov1984properties,moriya2014cubic,nakamura2012experimental,xiao2012coupled}. This field leads to the Rashba and Dresselhaus effects, which produce spin-split bands and momentum-dependent spin textures \cite{gmitra2016first,marchenko2012giant,stranks2018influence,bihlmayer2015focus,tao2017reversible,di2012electric}. Such spin-momentum locking underpins spin-selective transport, a key mechanism for spintronic device functionality.

The linear Rashba (LR) and linear Dresselhaus (LD) effects are particularly relevant in 2D systems. These effects are governed by SOC Hamiltonians:
\[
H_{LR} = \alpha_R (\sigma_x k_y - \sigma_y k_x), \quad
H_{LD} = \alpha_D (\sigma_x k_x - \sigma_y k_y),
\]
where \(\alpha_R\) and \(\alpha_D\) are the coupling strengths, and \(\sigma_i\) are Pauli matrices \cite{tao2017reversible}. These mechanisms result in energy dispersion \(E_k = \alpha k^2 \pm \tau k\), where \(\tau\) is the linear spin-splitting term. However, in systems governed by LR or LD effects, spin dephasing mechanisms such as Elliott–Yafet and Dyakonov–Perel make experimental realization of spin currents more challenging \cite{arras2019rashba,djani2019rationalizing,rosenblum2018progress,jia2020persistent,schliemann2003nonballistic}.

A particularly interesting solution to suppress spin dephasing is the emergence of a persistent spin texture (PST), where the spin orientation remains fixed regardless of the electron's momentum direction. PST occurs when LR and LD effects are balanced (\(\alpha_R = \alpha_D\)), leading to protected spin configurations and enabling long spin lifetimes \cite{schliemann2017colloquium}. Such balance can be engineered via structural tuning, gating, or carrier concentration control. While PSTs have been commonly associated with nonsymmorphic symmetries, recent theoretical and experimental efforts have shown that they can also arise in symmorphic space groups, provided certain mirror or rotational symmetries are present \cite{tao2018persistent}. These symmetries can constrain the spin-orbit field in a way that protects the spin configuration across momentum space, supporting nondissipative spin transport even in more conventional crystal symmetries.

In this work, we focus on 2D polar HOIPs that belong to the orthorhombic crystal family and exhibit spin-split states as a result of broken inversion symmetry and strong SOC. We demonstrate, through a combination of \(\mathbf{k} \cdot \mathbf{p}\) models and first-principles density functional theory (DFT) calculations, that PSTs can emerge in these systems—specifically along \(k\)-paths that are either symmetry-protected or lack mirror operations. We find that three representative compounds, \((\text{MIPA})_2\text{PbI}_4\), \((\text{MBPA})_2\text{PbBr}_4\), and \((\text{DMIPA})_2\text{PbI}_4\), satisfy the necessary symmetry and structural conditions to host PSTs. Our findings broaden the landscape of PST-hosting materials and open new possibilities for HOIP-based spintronic applications.

\begin{figure}[ht]
	\centering
	\includegraphics[width=\linewidth]{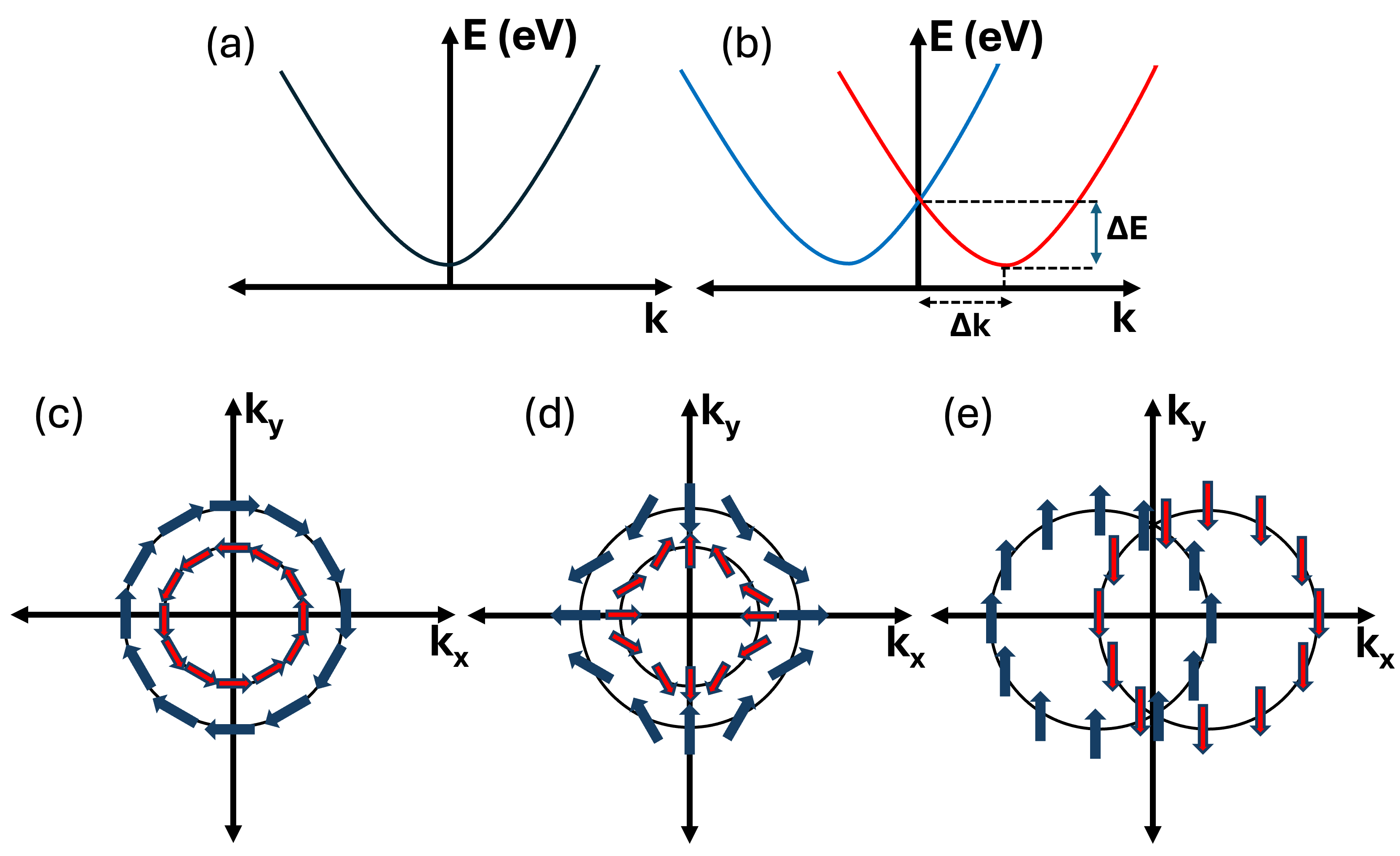}
	\caption{(a) Illustration of a doubly degenerate conduction band at \(k = 0\). 
		(b) Spin-split bands along a high-symmetry \(k\)-path, where \(\Delta E\) represents the splitting energy, and \(\Delta k\) denotes the momentum shift at the energy minimum. 
		(c), (d), and (e) depict schematics of the spin textures for Rashba, Dresselhaus, and persistent spin splitting mechanisms, respectively. 
		The blue and red arrows indicate the spin orientations in the outer and inner bands, respectively.}
	\label{fig:schematic_texture}
\end{figure}

\section{Methodology}
To explore the electronic and spin properties of our materials, we performed a series of comprehensive first-principles calculations using the Vienna ab initio simulation package (VASP) \cite{kresse1993ab,kresse1996computational}. These calculations are rooted in Density Functional Theory (DFT) \cite{hohenberg1964inhomogeneous,kohn1965self} and leverage the projector augmented wave (PAW) \cite{kresse1994norm,kresse1999ultrasoft} method for describing electron-ion interactions. For the exchange-correlation functional, we have used the Perdew-Burke-Ernzerhof revised for solids (PBEsol), which is widely applied for its balanced accuracy in predicting ground-state properties \cite{perdew2008restoring}. To further validate our results and achieve a more refined picture of the excited state properties, we employed the hybrid Heyd–Scuseria–Ernzerhof (HSE06) functional, known for its superior performance in electronic structure predictions \cite{heyd2003hybrid}. 

Throughout the calculations, we adopted a high energy cutoff of 600 eV, ensuring robust convergence and accuracy. The orthorhombic phase, denoted as Aea2 is fully optimized in terms of both lattice parameters and atomic positions, using a 4 × 4 × 4 k-point grid. The grid is generated using the Monkhorst–Pack scheme \cite{monkhorst1976special}, which effectively samples the Brillouin zone. To verify that spin-orbit coupling (SOC) has a negligible impact on structural relaxation, we performed additional test calculations, confirming the stability of our results. 

For electronic structure analysis, we computed the partial density of states (PDOS) and band structures using a finer 6 × 6 × 6 k-grid. The convergence criteria sets to be highly stringent: a total energy difference of less than $10^{-5}$ eV between consecutive ionic relaxation steps, and a force tolerance between atoms limited to 0.0001 eV.\AA. 

To capture the polarization within the material, we have calculated using Berry phase theory \cite{resta1994macroscopic,king1993theory,spaldin2012beginner}. To identify the symmetries, we have utilized tools such as the Bilbao Crystallographic Server \cite{aroyo2006krou,elcoro2017double}, SEEK-PATH \cite{hinuma2017band}, and Findsym \cite{stokes2005findsym}, which helped in a detailed symmetry analysis with ref. \cite{koster1963properties}. Additionally, Mathematica \cite{mathematica12} and PyProcar \cite{herath2020pyprocar} is used for visualizing the spin textures and solving the k.p model Hamiltonian, providing a clear representation of spin dynamics. 

For our spin texture calculations, the expectation values of spin operators $S_{i}$ were derived using the Pauli matrices and spinor eigenfunctions obtained from noncollinear spin calculations. This analysis was performed over a closely spaced $13 \times 13$ $\mathbf{k}$-grid around high symmetry points (HSPs) in the $k_x - k_y$ plane, covering a square region defined by $\left|k_x\right|$ and $\left|k_y\right| \leq 0.125 \, \text{\AA}^{-1}$. Such dense sampling ensures a detailed understanding of spin configurations around critical regions in momentum space.

To fit our $\mathbf{k} \cdot \mathbf{p}$ model to the DFT results, we used a Gaussian-like weight distribution centered around each high symmetry point (HSP), focusing on the range of $k_x$ and $k_y$ values from $-0.125 \, \text{\AA}^{-1}$ to $0.125 \, \text{\AA}^{-1}$. This approach minimizes discrepancies and provides an accurate representation of band structures near these symmetry points.

The model band structures and spin textures are obtained by parameterizing the models through the minimization of the summation

\[
S = \sum_{i=1}^{2} \sum_{\mathbf{k}} f(\mathbf{k}) \left| \det\left[ H(\mathbf{k}) - E^i(\mathbf{k}) I \right] \right|^2		\tag{3}
\]

 over the \(i\)-th energy eigenvalues \([E_i(\mathbf{k})]\) as training sets, where \(f(\mathbf{k})\) is the weight assigned to each \(\mathbf{k}\)-point. In this context, \(H\), ``Det,'' and \(I\) refer to the model Hamiltonian, the determinant, and the identity matrix, respectively. A normal distribution is used for \(f(\mathbf{k})\) to achieve a better fit near the high-symmetry points, as demonstrated in Refs. \cite{zhao2020large,zhao2020purely}.

Furthermore, we introduced strain and stress into the system to study their impact on electronic and spin properties. We performed the strain calculation by varying the $a$ and $b$-axis parameters relative to their equilibrium values ($a_0$, $b_0$). The corresponding strain ($\varepsilon$) was calculated using the formula $\varepsilon = \frac{b - b_0}{b_0}$, with the same approach for the $a$-axis, where \(b\) and \(b_0\) are replaced by \(a\) and \(a_0\). Both compressive (negative) and tensile (positive) strains were explored, ranging from $-5\%$ to $+5\%$. For the stress calculation, we applied pressure to our system ranging from $-3 \, \text{GPa}$ to $3 \, \text{GPa}$. After applying the strain or stress, we re-optimized the atomic coordinates to account for any structural relaxation effects, ensuring that the changes in spin textures and band structures are solely due to the applied strain or stress.

\section{Results and Discussions}
\begin{figure}[ht]
	\centering
	\includegraphics[width=\linewidth]{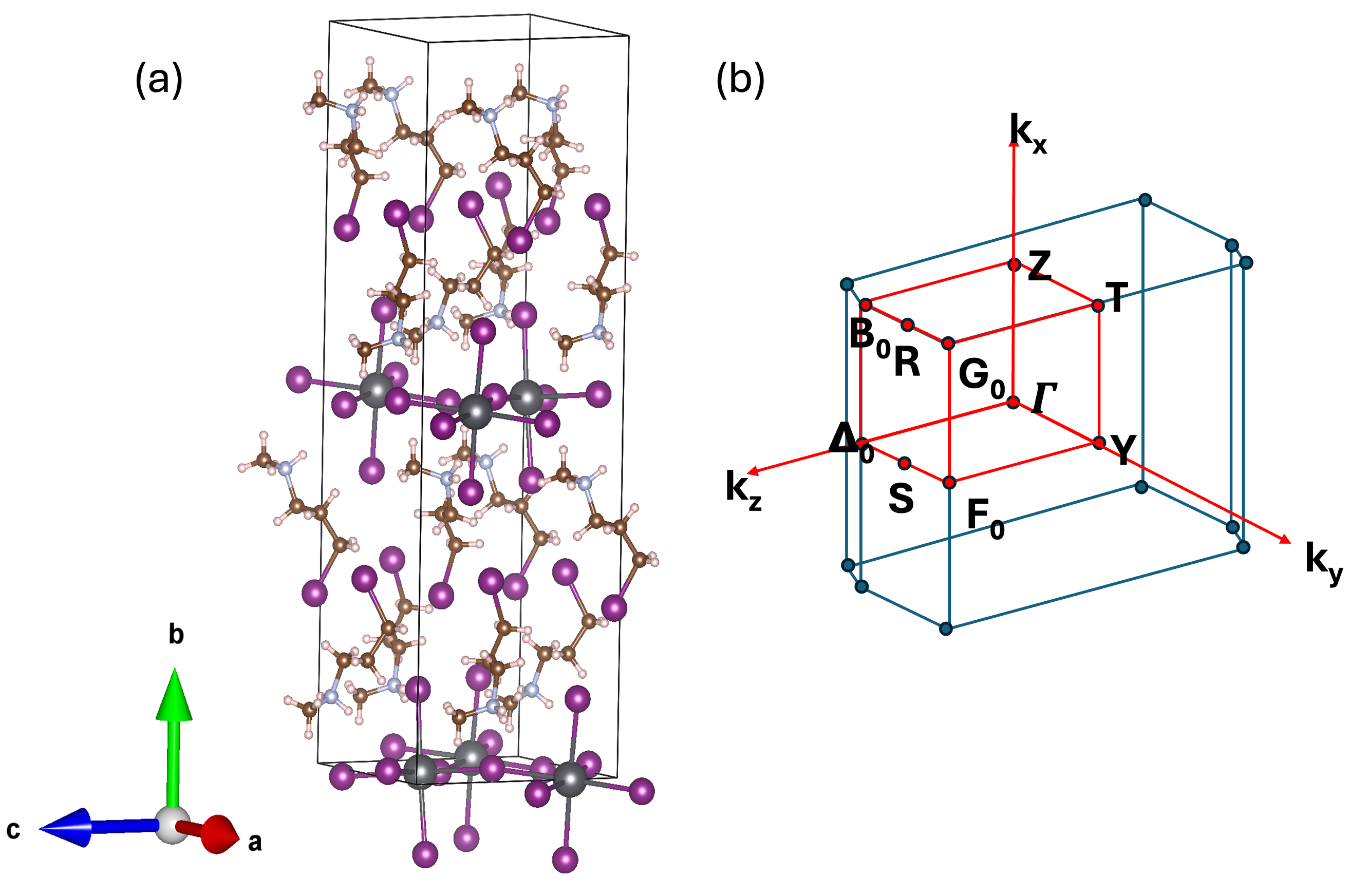}
	\caption{(a) Structure of \((\text{MIPA})_2\text{PbI}_4\) in Aea2 space group. The black, purple, grey, brown, and light pink balls denote the Pb, I, N, C and H, respectively. 
		(b) Brillouin Zone for respective materials including high symmetric \(k\)-path Y (0,$k_y$,0) - $\Gamma$ (0,0,0) - Z ($k_x$,0,0) is shown.}
	\label{fig:Structure}
\end{figure}

We have studied the experimentally synthesized family of layered organic-inorganic lead halide perovskites \cite{chakraborty2023rational}. This large family crystallizes in space group 41 (Aea2) and can exhibit persistent spin textures (PST). Specifically, we are interested in materials that are stable, have a wide band gap, and exhibit large spin splitting. Therefore, we have particularly focused on \((\text{MIPA})_2\text{PbI}_4\), \((\text{MBPA})_2\text{PbBr}_4\), and \((\text{DMIPA})_2\text{PbI}_4\). This family of materials possesses \(C_{2v}\) point group symmetry, which includes \(C_{2z}\) rotational, \(M_x\), and \(M_y\) glide plane symmetries. The relaxed crystal structure and atomic positions for \((\text{MIPA})_2\text{PbI}_4\) and Brillouin zone for the family are shown in Fig. \ref{fig:Structure}. Our relaxed crystal structures show good agreement with experimentally synthesized crystal structures. The calculated lattice parameters \((a, b, c)\) deviate from the experimental ones by \((0.71\%, 0.90\%, 0.61\%)\), \((1.01\%, 0.70\%, 0.66\%)\), and \((0.66\%, 0.76\%, 0.40\%)\) for \((\text{MIPA})_2\text{PbI}_4\), \((\text{MBPA})_2\text{PbBr}_4\), and \((\text{DMIPA})_2\text{PbI}_4\), respectively.
\begin{table*}
	\centering
	\renewcommand{\arraystretch}{1.5}
	\caption{Transformations of \((\sigma_x, \sigma_y, \sigma_z)\) and \((k_x, k_y, k_z)\) under the action of the generators of the \(C_{2v}\) point group and the time-reversal operator \(T\). The first row lists the point-group operations and their corresponding symmetries. Since the generators fully define the group structure, only these generators, along with the time-reversal operation \(T = i \sigma_y K\) (where \(K\) is the complex-conjugation operator), are used in constructing the \(k \cdot p\) model. The last row shows the terms that remain invariant under the point-group operations. Terms up to cubic order in \(k\) are included, as higher-order contributions are negligible.}
	\begin{tabular}{|>{\raggedright\arraybackslash}p{0.2\textwidth}|p{0.2\textwidth}|p{0.2\textwidth}|p{0.2\textwidth}|p{0.2\textwidth}|}
		\hline
		\textbf{Operations} & \(C_{2z} = e^{-i\pi / 2 \sigma_z}\) & \(M_{yz} = i \sigma_x\) & \(M_{xz} = i \sigma_y\) & \(T = i \sigma_y K\) \\ \hline
		\(k_x\) & \(-k_x\) & \(-k_x\) & \(k_x\) & \(-k_x\) \\ \hline
		\(k_y\) & \(- k_y\) & \(k_y\) & \(-k_y\) & \(-k_y\) \\ \hline
		\(k_z\) & \(k_z\) & \(k_z\) & \(k_z\) & \(-k_z\) \\ \hline
		\(\sigma_x\) & \(-\sigma_x\) & \(\sigma_x\) & \(-\sigma_x\) & \(-\sigma_x\) \\ \hline
		\(\sigma_y\) & \(-\sigma_y\) & \(-\sigma_y\) & \(\sigma_y\) & \(-\sigma_y\) \\ \hline
		\(\sigma_z\) & \(\sigma_z\) & \(-\sigma_z\) & \(-\sigma_z\) & \(-\sigma_z\) \\ \hline
		\textbf{Invariants} & 
		\(\begin{array}{l} 
			k_i^m k_x \sigma_x, k_i^m k_x \sigma_y, \\
			k_i^m k_y \sigma_x, k_i^m k_y \sigma_y \\
			k_i^m k_z \sigma_z \\
			(i = x, y, z; m=0,2)
		\end{array}\)
		&
		\(\begin{array}{l} 
			k_i^m k_x \sigma_y, k_i^m k_y \sigma_x, \\
			k_i^m k_x \sigma_z, k_i^m k_z \sigma_x \\
			(i = x, y, z; m=0,2)
		\end{array}\)
		&
		\(\begin{array}{l} 
			k_i^m k_x \sigma_y, k_i^m k_y \sigma_x, \\
			k_i^m k_y \sigma_z, k_i^m k_z \sigma_y \\
			(i = x, y, z; m=0,2)
		\end{array}\)
		&
		\(\begin{array}{l} 
			k_i \sigma_j \\
			(i,j = x, y, z)
		\end{array}\)
		\\ \hline
	\end{tabular}
\end{table*}

\begin{figure}[H] 
	\centering
	\includegraphics[width=\linewidth, height=0.55\textheight]{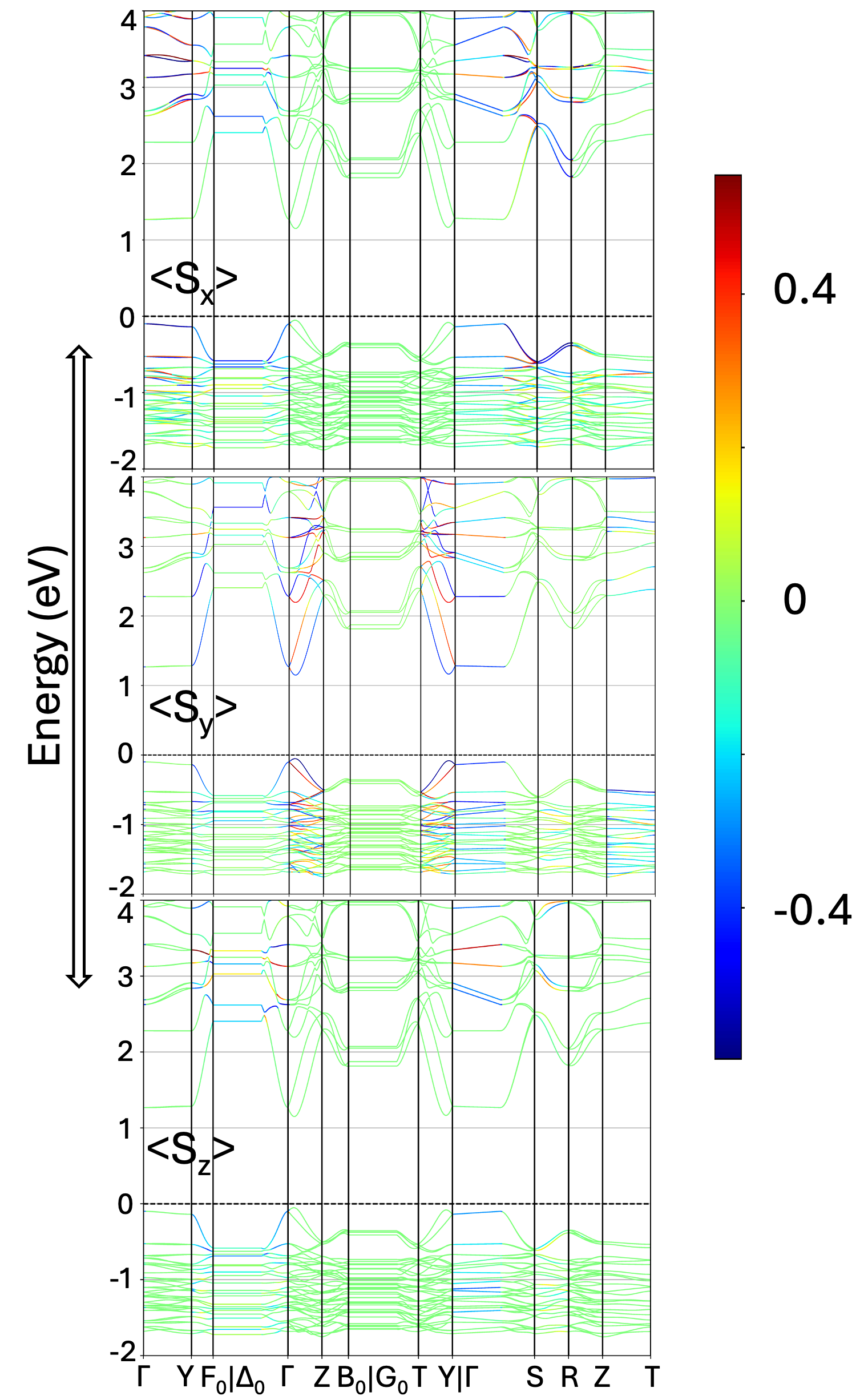} 
	\caption{ The spin-resolved band structures along the high-symmetry path are depicted. The color bars indicate the expectation values of the spin components $S_x$, $S_y$, and $S_z$.}
	\label{fig:Band_Structure}
\end{figure}

We have calculated the band structures using the PBEsol exchange-correlation functional, with and without the inclusion of spin-orbit coupling effects. Fig. \ref{fig:Band_Structure} shows the spin-projected band structure of \((\text{MIPA})_2\text{PbI}_4\). A band gap of \(1.20 \, \text{eV}\) is calculated with the conduction band minimum (CBm) and valence band maximum (VBM) at the \(\Gamma\) and \(Z\) points, respectively. It is known that PBEsol underestimates the band gap; therefore, we have also calculated the band gap using the hybrid exchange-correlation functional HSE06. We have obtained a band gap of \(2.07 \, \text{eV}\) for \((\text{MIPA})_2\text{PbI}_4\). Similarly, we have done the calculation of band structures for \((\text{MBPA})_2\text{PbBr}_4\) and \((\text{DMIPA})_2\text{PbI}_4\) [See SM].

Incorporating the effect of spin-orbit coupling (SOC) leads to the splitting of bands throughout the Brillouin zone (BZ). In \((\text{MIPA})_2\text{PbI}_4\), spin splitting is present at the valence band maximum (VBM) and conduction band minimum (CBm) along the \(\Gamma\) to \(Z\) path. This is due to the contribution of Pb-5d and I-4p orbitals in the upper valence band and lower conduction band. Along the \(\Gamma\)-\(Z\) path, it is clearly visible that the band dispersion is primarily characterized by the \(y\)-component of spin, whereas the \(x\)- and \(z\)-components of spin are absent along that path. The in-plane mirror symmetry of the crystal \((M_{xz})\) enforces the relation \((S_x, S_y, S_z) \rightarrow (-S_x, S_y, -S_z)\), which is satisfied only when \(S_x\) and \(S_z\) are zero. This leads to the spin-orbit field aligning along the \(y\)-axis.

\begin{figure*}[ht]
	\centering
	\includegraphics[width=\linewidth]{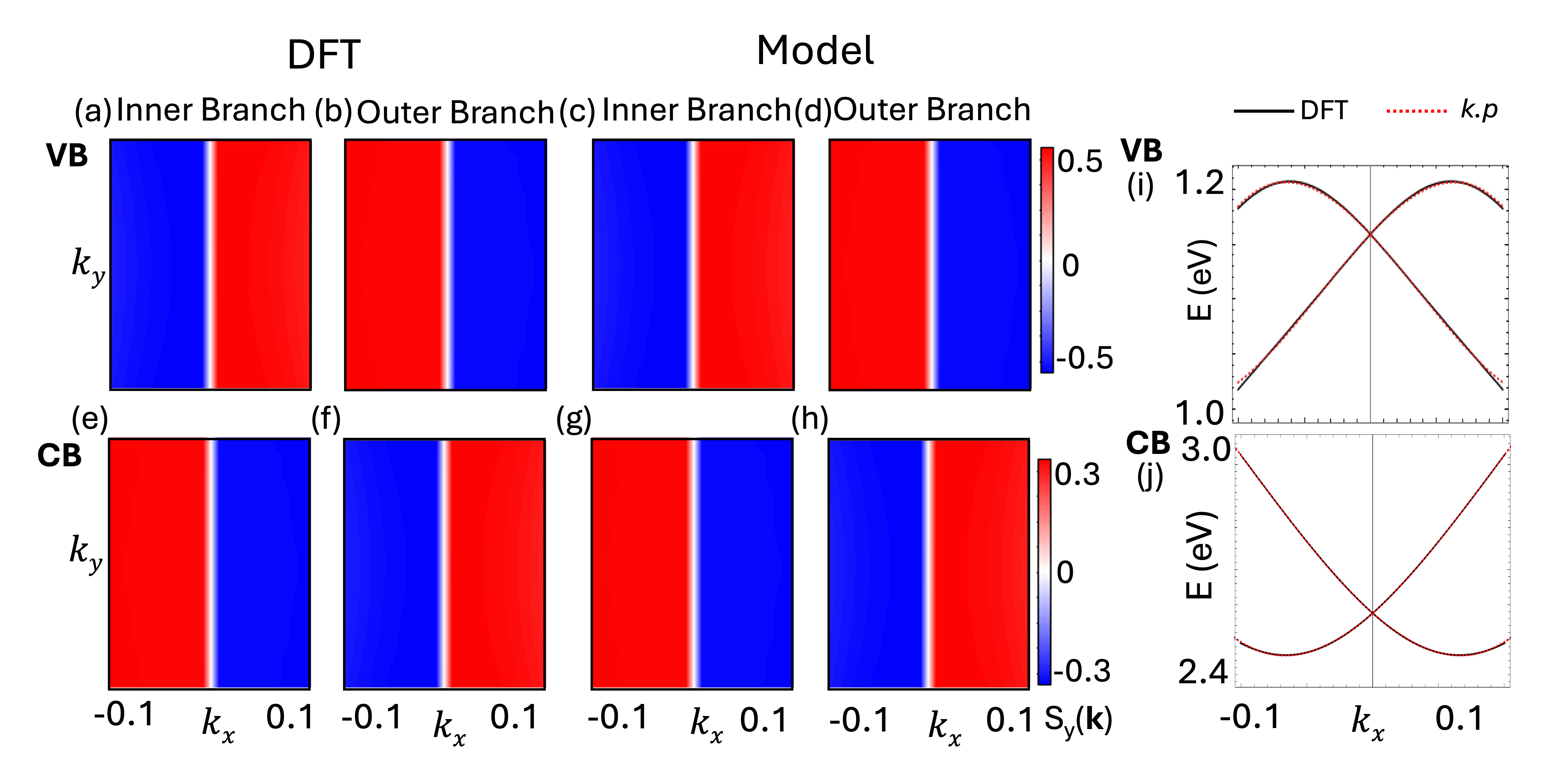}
	\caption{Spin textures of the valence band for (MIPA)$_2$PbI$_4$ around the $\Gamma$ point obtained using DFT [(a), (b)] and the $k \cdot p$ model [(c), (d)], and spin textures of the conduction band for (MIPA)$_2$PbI$_4$ around the $\Gamma$ point obtained using DFT [(e), (f)] and the $k \cdot p$ model [(g), (h)]. Spin textures are presented using the convention [(a)–(h)] where the energy of the inner branch is less than or equal to that of the outer branch at any $(k_x, k_y)$ point in the given range. The color represents the $y$ component of the spin textures. The $x$ component shows a negligible contribution compared to the $y$ component around the $\Gamma$ line, while the $z$ component is zero in the $k_x$–$k_y$ plane. (i) Band structure of the valence band maximum (VBM) and (j) conduction band minimum (CBm) of (MIPA)$_2$PbI$_4$ around the $\Gamma$–Z line (energy vs. $k_x$), projected onto the $y$ component of the spin direction. The black solid lines and red dashed lines represent results obtained using DFT and model parameterization, respectively.}
	\label{fig:Spin_Texture}
\end{figure*}

The Fig. \ref{fig:Spin_Texture} shows the spin splitting at the VBM and CBm, along with the corresponding spin textures. It is observed that the spin-split bands are predominantly linear in \(\mathbf{k}\). For a more comprehensive understanding, the band dispersion can be derived by identifying all symmetry-allowed terms, such that

\[
H(\mathbf{k}) = O^\dagger H(\mathbf{k}) O,
\]

where \(O\) represents the symmetry operations corresponding to the wave vector group (\(G\)) associated with the high-symmetry point, along with time-reversal symmetry (\(T\)) \cite{voon2009kp}. The Hamiltonian that remains invariant under these operations satisfies the following condition:

\[
H_G(\mathbf{k}) = D(O) H(O^{-1} \mathbf{k}) D^{-1}(O), \quad \forall O \in G, \, T,
\]

where \(D(O)\) is the matrix representation of the operation \(O\) belonging to the point group \(G\). The \(\mathbf{k} \cdot \mathbf{p}\) Hamiltonian is derived using the method of invariants by considering the little group of the \(\Gamma\)-point to be \(C_{2v}\), which includes glide planes \(M_{xz}\) and \(M_{yz}\), along with the rotational symmetry \(C_{2z}\) and the identity operation (\(E\)). The \(\mathbf{k} \cdot \mathbf{p}\) Hamiltonian around the \(\Gamma\)-point, following the transformation rules outlined in Table I, is given by \cite{kumar2023exploring}

\[
H_{C_{2v}}(\mathbf{k}) = H_0(\mathbf{k}) + \alpha(\mathbf{k}) k_x \sigma_y + \beta(\mathbf{k}) k_y \sigma_x,		\tag{4}
\]

where \(k_x\) and \(k_y\) are the Cartesian components of the crystal momentum in the plane around the \(\Gamma\)-point, and \(\boldsymbol{\sigma}\) are the Pauli matrices that describe the spin degrees of freedom. \(H_0(\mathbf{k})\) represents the part of the Hamiltonian that describes the band dispersion, which depends on the parameters \(\delta\) and \(\eta\), as

\[
H_0(\mathbf{k}) = E_0 + \delta k_x^2 + \eta k_y^2.			\tag{5}
\]

Here, \(\delta\) and \(\eta\) are related to the effective masses \(m_x^*\) and \(m_y^*\) by the relations

\[
|\delta| = \frac{\hbar^2}{2m_x^*}, \quad |\eta| = \frac{\hbar^2}{2m_y^*}.		
\]

The \(k\)-dependent spin-orbit coupling constants in \(H_{C_{2v}}(\mathbf{k})\) are given by

\[
\alpha(\mathbf{k}) = \alpha^{(1)} + \alpha^{(3)} k^2 + \gamma_\alpha (k_x^2 - k_y^2),	\tag{6}
\]

\[
\beta(\mathbf{k}) = \beta^{(1)} + \beta^{(3)} k^2 + \gamma_\beta (k_x^2 - k_y^2),		\tag{7}
\]

where \(\alpha^{(3)}\) and \(\beta^{(3)}\) represent the \(k^2\)-dependent renormalization terms for the linear spin-orbit coupling constants, while \(\gamma_\alpha\) and \(\gamma_\beta\) account for the \(k\)-cubic anisotropic interactions.

The Eigenstates corresponding to Hamiltonian $H_{C_{2v}}(\mathbf{k})$ are

\[
\psi_{\mathbf{k}, \uparrow} = \frac{e^{i \mathbf{k} \cdot \mathbf{r}}}{\sqrt{2 \pi}}
\begin{pmatrix}
	-\frac{A - i B}{E_{SO}} \\
	1
\end{pmatrix}		\tag{8}
\]
and 
\[
\psi_{\mathbf{k}, \downarrow} = \frac{e^{i \mathbf{k} \cdot \mathbf{r}}}{\sqrt{2 \pi}}
\begin{pmatrix}
	\frac{A - i B}{E_{SO}} \\
	1
\end{pmatrix}		\tag{9}
\]
\text{where}
\[
A = \beta k_y + \delta \left(k_y^3 + k_x^2 k_y\right),
\quad
B = \alpha k_x + \gamma \left(k_x^3 + k_x k_y^3\right),
\]
\[\quad \text{and} \quad
E_{SO} = \sqrt{A^2 + B^2}.
\]

The corresponding eigenvalues are given as 
\[
E_{C_{2v}}(\mathbf{k}) = E_0(\mathbf{k}) \pm E_{SO}		\tag{10}
\]
The spin textures, determined by the expression 
$\mathbf{S}_\pm = \frac{\hbar}{2} \langle \psi_{\mathbf{k}\uparrow,\downarrow} | \boldsymbol{\sigma} | \psi_{\mathbf{k}\uparrow,\downarrow} \rangle$ are given as

\begin{align}
	\langle \sigma_x^{\pm} \rangle &= \pm \frac{\beta(k) k_y}{\sqrt{\alpha(k)^2 k_x^2 + \beta(k)^2 k_y^2}}, \tag{11} \\
	\langle \sigma_y^{\pm} \rangle &= \pm \frac{\alpha(k) k_x}{\sqrt{\alpha(k)^2 k_x^2 + \beta(k)^2 k_y^2}}, \tag{12} \\
	\langle \sigma_z^{\pm} \rangle &= \pm 0. \tag{13}
\end{align}

\begin{table*}[t]
	\centering
	\renewcommand{\arraystretch}{1.5}
	\label{tab:Spin}
	\caption{The materials showing PST along with their parameters of Eq.~(4). The splitting is observed for the conduction 
		band minimum (CBm) and valence band maximum (VBM) around the high-symmetry points (HSP) $\Gamma$ to Z.}
	\begin{tabular}{|>{\raggedright\arraybackslash}p{0.13\textwidth}|p{0.09\textwidth}|p{0.08\textwidth}|p{0.08\textwidth}|p{0.09\textwidth}|p{0.09\textwidth}|p{0.1\textwidth}|p{0.1\textwidth}|p{0.09\textwidth}|p{0.09\textwidth}|}
		\hline
		\textbf{Materials} & VBM/CBm & $\delta$ (eV·\AA$^2$) & $\eta$ (eV·\AA$^2$) & $\alpha^{(1)}$ (eV·\AA) & $\beta^{(1)}$ (eV·\AA) & $\alpha^{(3)}$ (eV·\AA$^3$)& $\beta^{(3)}$ (eV·\AA$^3$) & $\gamma_\alpha$ (eV·\AA$^3$) & $\gamma_\beta$ (eV·\AA$^3$) \\ \hline
		
		(\text{MIPA})$_2$PbI$_4$ & VBM  & -5.41 & -1.88 & 1.28 & 0.01 & -4.52 & 0.23 & -43.68 & 0.46 \\ \hline
								 & CBm  & 18.39 & 0.79 & 3.33 & 0.04 & -7.18 & 0.98 & -59.04 & 0.37 \\ \hline
		(\text{MBPA})$_2$PbBr$_4$ & VBM  & -4.78 & -1.32 & 0.73 & 0.01 & -10.15 & 0.15 & -11.00 & -0.18 \\ \hline
								 & CBm  & 13.64 & 0.50 & 3.12 & 0.01 & -6.52 & 0.31 &	-30.43 & 0.10 \\ \hline
		(\text{DMIPA})$_2$PbI$_4$ & VBM  & -17.98 & -0.31 & 0.82 & 0.01 & -6.86 & 0.15 & -12.33 & -0.06 \\ \hline
								 & CBm  & 27.06 & 0.60 & 1.20 & 0.02 & -3.27 & 0.43 & -29.46 & 0.32 \\ \hline
		(\text{MIPA})$_2$PbBr$_4$ & VBM  & -4.86 & -1.30 & 0.65 & 0.01 & -0.76 & 0.15 & -9.75 & 0.26 \\ \hline
								 & CBm  & 11.97 & 0.55 & 1.18 & 0.02 & -1.22 & 0.68 & -10.88 & 0.17 \\ \hline
		(\text{MIPA})$_2$SnI$_4$ & VBM  & -17.96 & -1.52 & 0.86 & 0.01 & -13.59 & 0.18 & -19.77 & 0.41 \\ \hline
							 & CBm  & 18.45 & 0.48 & 3.12 & 0.03 & -23.04 & 0.76 & -67.14 & 0.28 \\ \hline
		(\text{MBPA})$_2$SnBr$_4$ & VBM  & -13.18 & -0.97 & 0.66 & 0.01 & -2.31 & 0.12 & -11.69 & -0.11 \\ \hline
								 & CBm  & 14.33 & 0.27 & 2.04 & 0.01 & -12.30 & 0.21 & -41.11 & 0.06 \\ \hline
		(\text{MBPA})$_2$PbI$_4$ & VBM  & -3.71 & -1.40 & 1.67 & 0.01 & -5.01 & 0.42 & -46.40 & -0.34 \\ \hline
								 & CBm  & 14.64 & 0.96 & 4.55 & 0.01 & -10.53 & 0.75 & -80.39 & 0.25 \\ \hline
		(\text{DMIPA})$_2$SnI$_4$ & VBM & -15.57 & -0.25 & 1.34 & 0.01 & -16.29 & 0.12 & -10.17 & -0.04 \\ \hline
								 & CBm  & 17.95 & 0.42 & 1.84 & 0.01 & -9.10 & 0.36 & -70.91 & 0.27 \\ \hline
	\end{tabular}
\end{table*}

Equation (11), (12) and (13) show spin component $\sigma_x$, $\sigma_y$ and $\sigma_z$ are present around the $\Gamma$ point and is found to be consistent with DFT spin texture shown in Fig. \ref{fig:Spin_Texture}(c), \ref{fig:Spin_Texture}(d), \ref{fig:Spin_Texture}(g), and \ref{fig:Spin_Texture}(h). For \((\text{MIPA})_2\text{PbI}_4\), By fitting the band structure and spin texture around $\Gamma$ point using both DFT and the model hamiltonian, we find that $\alpha^{(1)}$ = 3.6 eV·Å for CBm and $\alpha^{(1)}$ = 1.28 eV·Å for VBM. This splitting is significantly larger as compared to other bulk systems such as in MAPbI\(_3\) ($\alpha$ = 2.50 eV·Å), (\text{BA})$_2$PbCl$_4$ ($\alpha$ = 0.69 eV·Å), and (\text{FPEA})$_2$PbI$_4$ ($\alpha$ = 0.36 eV·Å). It is also observed that around $\Gamma$ point, $\beta^{(1)}$ is nearly zero, compare to $\alpha^{(1)}$. The linear spin splitting coefficient $\alpha^{(1)}$ is zero or one order of magnitude smaller than the cubic terms $\alpha^{(3)}$ and $\gamma_\alpha$ and on the other hand $\beta^{(1)}$ is one order smaller than cubic terms $\beta^{(3)}$ and $\gamma_\beta$. As long as $|\alpha^{(1)}| \gg |\alpha^{(3)}k^2|$, there is neglegible contributions of higher order coefficients, refers to Table II. By this observation, we can say that $\alpha_R$ is approximately equals to $\alpha_D$, as \[
\alpha_{R/D} = \frac{\alpha^{(1)} \pm \beta^{(1)}}{2} \approx \frac{\alpha^{(1)}}{2}
\] and this is the required condition for Persistent type spin texture. The orbital contribution also influences whether significant PST splitting occurs in the VB or CB. Substantial splitting happens when heavy-atom orbitals are involved in the splitting bands. For example, in \((\text{MIPA})_2\text{PbI}_4\), splitting is large in CB, due to the contribution of Pb and I orbitals.

Building on the confirmation of Persistent Spin Texture (PST) in \((\text{MIPA})_2\text{PbI}_4\), we next explore how external factors, such as strain and stress, can be utilized to fine-tune the spin splitting in two-dimensional (2D) layered perovskites. Strain engineering has emerged as a promising approach to modify the electronic structure in these materials. Experimental observations indicate that the electronic structure and spin splitting in two-dimensional (2D) layered perovskites are tunable using strain \cite{cheng2022impact}. The spin splitting is sensitive to uniaxial strain in layered perovskites such as MAPbI\(_3\), FAPbI\(_3\) \cite{mahajabin2024role}, MPSnBr\(_3\) \cite{kashikar2023persistent}, and others. In view of this, we have studied the strain dependence of the spin splitting. We considered the effect of strain along the \(a\)- or \(b\)-axis on spin splitting and energy levels for both phases around the VBM and CBm. Such strain can be induced by applying uniaxial stress or by growing \((\text{MIPA})_2\text{PbI}_4\) films on suitable substrates, allowing for controlled deformation along these crystallographic directions.

\begin{figure*}[t] 
	\centering
	\includegraphics[width=\textwidth, height=0.22\textheight]{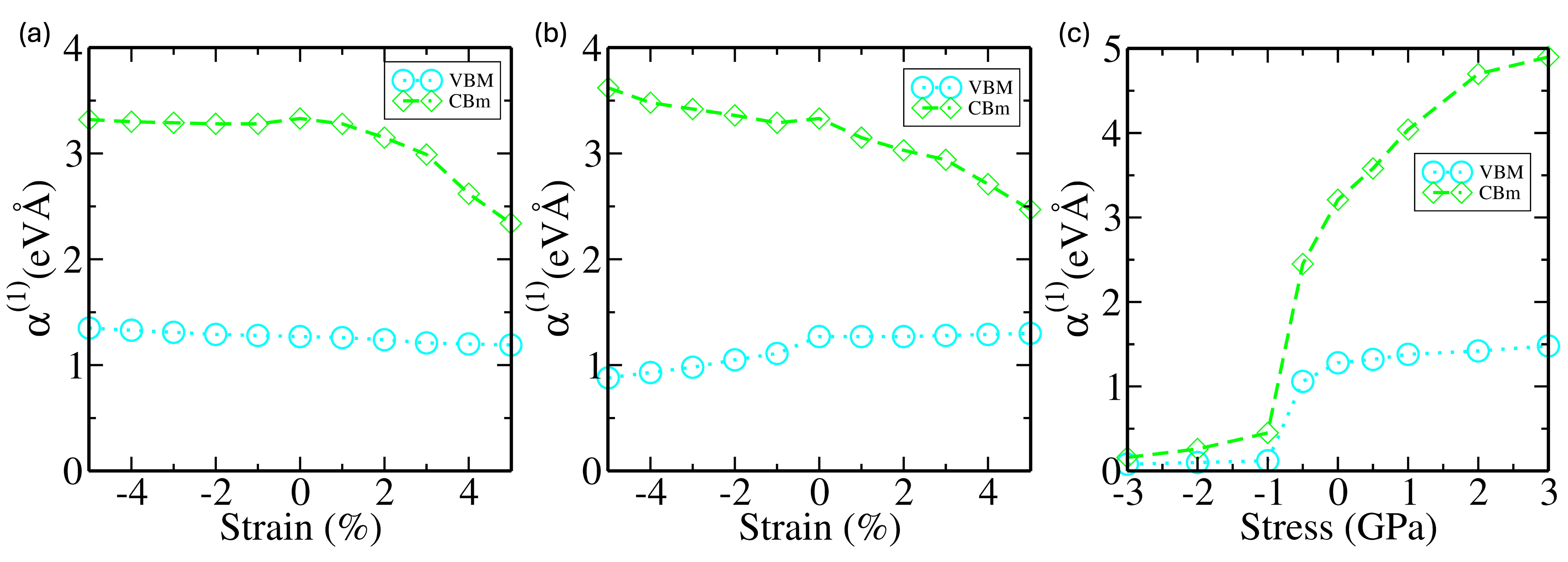}
	\caption{(a) and (b) show the variation in the linear spin splitting coefficient as a function of uniaxial strain in the \(a\)- and \(b\)-directions, respectively. (c) shows the change in the spin splitting coefficient as a function of stress (GPa).}
	\label{fig:Strain}
\end{figure*}

Fig. 5(a) and 5(b) shows the change in Rashba coefficient \(\alpha_R\) upon the application of strain. It is observed that for \((\text{MIPA})_2\text{PbI}_4\), the spin splitting in the CBm decreases with tensile strain and increases with compressive strain for both \(a\)- and \(b\)-axis strains, whereas for the VBM, the strain has a minimal effect. Specifically, for \(a\)-axis strain, spin splitting increases (decreases) with compressive (tensile) strain, while for \(b\)-axis strain, spin splitting increases (decreases) with tensile (compressive) strain. 

By applying strain from -5\% to +5\%, For \(a\)-axis strain, the linear spin splitting $\alpha^{(1)}$ for the CBm and VBM varies from 3.32 eV·Å to 1.90 eV·Å and from 1.37 eV·Å to 1.19 eV·Å, respectively. For \(b\)-axis strain, the spin splitting amplitude for the CBm and VBM varies from 3.70 eV·Å to 2.30 eV·Å and from 1.30 eV·Å to 0.77 eV·Å, respectively. We have calculated $\alpha^{(1)}$ vs strain (\%) for other family of materials also [See SM].

By applying strain, the electronic band structure can be tailored to enhance spin splitting conditions. This modification enables easier access to spin-polarized states at reduced energy levels, making the material more efficient for spintronic applications that demand lower power consumption. Strain was applied along the \(a\)- and \(b\)-crystallographic axes to explore directional control of spin splitting. Although the strain altered the spin splitting, the spin texture remained largely unaffected, indicating the stability of the material's spin properties under these conditions. This suggests that the spin characteristics can be optimized without significant changes in texture, providing flexibility in the design of spintronic devices.

To explore the tunability of spin splitting in response to applied stress within the elastic regime, we apply stress in the \((\text{MIPA})_2\text{PbI}_4\) system, ranging from \(-3\) GPa to \(+3\) GPa, to compute the band structure along the Z–\(\Gamma\)–Z direction. Here, positive stress represents tensile stress, while negative stress denotes compressive stress.
The linear spin-splitting coefficient is observed to vary with the application of compressive or tensile stress. For the VBM and CBm, \(\alpha^{(1)}\) ranges from 0.08~eV·\AA{} to 1.48~eV·\AA{} and from 0.16~eV·\AA{} to 4.9~eV·\AA{}, respectively, under stresses ranging from \(-3~\text{GPa}\) to \(3~\text{GPa}\). The evolution of spin splitting in both the conduction band (CB) and valence band (VB) as a function of applied stress is illustrated in Fig. 5(c), which demonstrates the high tunability of spin splitting in response to stress.

Our results indicate that spin splitting in the Z–\(\Gamma\)–Z direction increases under tensile stress and decreases under compressive stress, nearly vanishing at high compressive values. Furthermore, we examine the effect of stress on the spin texture and observe that the persistent spin texture characteristic is maintained across the elastic stress range. This finding highlights the ability to tune spin splitting without significant alteration of the spin texture, underscoring the potential of \((\text{MIPA})_2\text{PbI}_4\) in spintronic applications.

\section{Conclusion}
Combining first-principles calculations and symmetry analysis, we have shown the existence of PST in pseudo 2D materials having \(C_{2v}\) point-group symmetry. The unidirectional spin textures are observed in full k$_x$-k$_y$ plane, as long as in-plane mirror symmetry remains intact and differs from trivial PST. Taking \((\text{MIPA})_2\text{PbI}_4\) as a test case, we have observed linear spin splitting strength around the \(\Gamma\) point of an order 3.70 eV·Å. The observed splitting is found to be anisotropic in nature. The nature and anisotropy of splitting are also studied using the k · p model via symmetry analysis. Our calculations show that there exists a large family of materials which possess full-plane PST. In addition, strain engineering tunes the observed PST by varying the magnitudes of SOC splitting coefficients. These spin textures known for nondissipative spin transport, form another prospective aspect. The large SOC splitting coefficients, wide band gap, suitable band-edge positions, strain tunability, and room-temperature stability make them suitable for room-temperature applications. The complete realization of original device of these applications may enrich the field of spintronics.

\section{Acknowledgement}
S.P. acknowledges PMRF, India, for the Research fellowship [Grant No. 1403227]. S. B. acknowledges financial support from SERB under a core research grant (grant no. CRG/2019/000647) to set up his High Performance Computing (HPC) facility “Veena” at IIT Delhi for computational resources. We acknowledge the High Performance Computing (HPC) facility at IIT Delhi for computational resources.

\bibliography{ref}

\end{document}